\documentclass[]{spie}
\usepackage[]{graphicx}

\newcommand{\eg}{{\it e.g.}}

\title{Methods of optimizing X-ray optical prescriptions for wide-field 
applications}

\author{Ronald F. Elsner\supit{a}, Stephen L. O'Dell\supit{a}, Brian D. Ramsey\supit{a}, and Martin C. Weisskopf\supit{a}
\skiplinehalf
\supit{a}NASA Marshall Space Flight Center, Space Science Office, VP62, Huntsville, AL 35812
}

\authorinfo{Further author information: (Send correspondence to R.F.E)\\R.F.E.: E-mail: ron.elsner@nasa.gov, Telephone: 256 961 7765\\  S.L.O.: E-mail: steve.o'dell@nasa.gov, Telephone: 256 961 7776\\  B.D.R: E-mail: brian.ramsey@nasa.gov, Telephone: 256 961 7784\\  M.C.W.: Email: martin@smoker.msfc.nasa.gov, Telephone: 256 961 7798}

\begin{document}

\maketitle

\begin{abstract}
We are working on the development of  a method for optimizing wide-field X-ray telescope mirror prescriptions, including polynomial coefficients, mirror shell relative displacements, and (assuming 4 focal plane detectors) detector placement along the optical axis and detector tilt. 
With our methods, we hope to reduce number of Monte-Carlo ray traces required to search the multi-dimensional design parameter space, and to lessen the complexity of finding the optimum design parameters in that space.
Regarding higher order polynomial terms as small perturbations of an underlying Wolter I optic design, we begin by using the results of Monte-Carlo ray traces to devise trial analytic functions, for an individual Wolter I mirror shell, that can be used to represent the spatial resolution on an arbitrary focal surface.
We then introduce a notation and tools for Monte-Carlo ray tracing of a polynomial mirror shell prescription which permits the polynomial coefficients to remain symbolic.
In principle, given a set of parameters defining the underlying Wolter I optics, a single set of Monte-Carlo ray traces are then sufficient to determine the polymonial coefficients through the solution of a large set of linear equations in the symbolic coefficients.
We describe the present status of this development effort.

\end {abstract}



\section{Introduction}
\label{s:intro}  

In 1992, Burrows, Burg, and Giacconi\cite{BBG92} showed how by adding higher order polynomial terms to Wolter I prescriptions, and hence giving up some on-axis spatial resolution, one can obtain prescriptions for reflecting surfaces that provide improved average spatial resolution over a wide field-of-view (say $\sim 30$ arcmin).
Such so-called polynomial optics would be particularly useful for moderately deep to deep surveys, to be carried out by observatories such as for the proposed Wide-Field X-ray Telescope (WFXT) mission\cite{SSM08}, and for solar X-ray observations.
Procedures for optimizing the design of wide-field X-ray telescopes utilize Monte-Carlo methods for determining the design parameters, including specification of the polynomial coefficients\cite{BBG92,Roming00,Roming02,Conconi04,Conconi09,Conconi10}.
Monte-Carlo ray traces are performed over a range of design parameters, and the final design determined according to some optimization criterion and methods.
Since the number of mirror shells per module is typically large ($\sim$ 50---100), these procedures are presently complicated and computer intensive.

The present paper is a report on the current status of an on-going study\cite{RFE09} of the properties of Wolter I and polynomial optical prescriptions with the ultimate goal of simplifying the procedures for optimizing their designs.
Since a polynomial prescription can typically be viewed as a small pertubation to an underlying Wolter I design, we begin with Monte-Carlo studies of the properties of Wolter I mirror shells relevant to wide-field designs, attempting to deduce analytic formulae for representing the geometric area and spatial resolution as functions of source position on the sky relative to the pointing axis, focal length, mirror shell segment length and shell intersection radius.
We then outline a method, valid when the polynomial coefficients are sufficiently small, for ray tracing polynomial optics keeping the polynomial coefficients in symbolic form.

A merit function providing a measure of spatial resolution averaged over the field-of-view (FOV) is defined in \S\ref{s:merit}, while the parameter space over which we have carried out Monte-Carlo ray traces is described in \S\ref{s:montecarlo}.
We note in \S\ref{s:single1} that the spatial resolution, when averaged over the FOV as in the merit function, as a function of source position relative to the optical axis, is a simple sum of terms up to second order in (1) the mirror shell displacement relative to the nominal on-axis focus, and (2) the tilt angle for the CCD detector array.
In \S\ref{s:nested} we arrive at the important conclusion that the spatial resolution on an arbitrary focal surface for a set of nested mirror shells may be written as the sum of two terms.
The first is a sum over the spatial resolution of the individual shells on that surface, weighted by their effective area.
The second is a sum over a kind of weighted variance of the mean ray positions for the individual telescopes on that surface.
In \S\ref{s:polynotation}, we introduce a compact notation for representing ray trace variables such as position or direction vectors, including polynomial coefficients in symbolic form.
In \S\ref{s:outer}, we discuss the outer product of two vectors, a concept from linear algebra necessary for the development of the methods introduced in this paper.
In \S\ref{s:basic}, we specify the basic operations of a polynomial optic algebra, which are addition, subtraction, multiplication, division, and the taking of square roots.
Given a direction vector, $\vec{k_1}$, and initial position $\vec{x_1} = ( x_1, y_1, z_1)$, \S\ref{s:rayprop} shows how to propagate a ray from $\vec{x_1}$ to axial position $z_2$ and determine the other coordinates, $x_2$ and $y_2$, thus determining the final position $\vec{x_2}$.
We define our coordinate system and the mirror surface prescriptions for polynomial X-ray optics in \S\ref{s:poly}.
In \S\ref{s:polyraytracing}, we list the tasks required to trace rays through X-ray optics.
In the future, we plan to show how the tools presented in this paper are used to accomplish these tasks while keeping the polymonial coefficients in symbolic form, and to provide concrete examples.
Some closing remarks are provided in \S\ref{s:remarks}.


\section {Merit function}
\label{s:merit}

For X-ray survey applications, such as the proposed Wide-Field X-ray Telescope 
(WFXT) mission\cite{SSM08}, one desires a large effective collecting area over a broad energy range combined with good spatial resolution over a wide FOV.
The geometric area available is essentially pre-determined by the diameter of the launch vehicle faring, the number of desired telescope modules (which are constrained by the desired FOV and, in the absence of extendable optical benches, the focal lengths permitted by the launch vehicle faring), and the number of mirror shells per module allowed by mass and manufacturing constraints.
In our work, we have therefore concentrated on optimizing the spatial resolution average over the FOV, by minimizing the merit function:

\begin {equation}
 \label{e:merit}
 M \ \equiv \ \frac{\int_{\phi=0}^{2 \pi} \ d \phi \ \int_{\theta=0}^{\theta_{FOV}} \theta \ d \theta \ w(\theta, \phi) \ \sigma^2(\theta, \phi)}{\int_{\phi=0}^{2 \pi} \ d \phi \ \int_{\theta=0}^{\theta_{FOV}} \theta \ d \theta \ w(\theta, \phi)},
\end {equation}

\noindent
where $\theta$ is the polar off-axis angle for the incident X-rays, $\phi$ is the azimuthal angle for the incident X-rays, and $w(\theta, \phi)$ is a weighting factor.
By symmetry, the average in Eq. (\ref{e:merit}) may be restricted to $\phi \in [0, \pi / 4]$ for a typical detector setup consisting of four tilted CCDs, each occupying a single quadrant.
This statement neglects any repositioning of the detectors to place the on-axis aim point on one of them.
The quantity $\sigma^2(\theta,\phi)$ is the variance in the position of rays reaching the focal surface.
This focal surface may be curved or tilted with respect to the flat plane perpendicular to the optical axis and passing through the nominal on-axis best focus.
The variance, $\sigma^2(\theta,\phi)$, is given by

\begin {equation}
 \label{e:sigmadefn}
 \sigma^2(\theta,\phi) \ = \ [ \ ( \ < x^2 > \ - \ < x >^2 \ ) \ + \ ( \ < y^2 > \ - \ < y >^2 \ ) \ + \ ( \ < z^2 > \ - \ < z >^2 \ ) \ ]
\end {equation}

\noindent where $x$, $y$ and $z$ are the positions of the rays on the chosen focal surface, and $< q >$ denotes an average of the quantity $q$.
All rays incident on the detector are included in the averages in Eq. (\ref{e:sigmadefn}), independent of the mirror shell from which they exited.

We have found that the coefficients of the polynomial terms modifying Wolter I optics for wide-field applications may be regarded as small for our purposes.
Therefore, we treat them as small perturbations to the underlying Wolter I design.
It is for this reason that we have sought analytical fitting functions for the contributions to $\sigma^2(\theta,\phi)$ for Wolter I optics.
This is also the justification for the procedure we describe later in this paper for ray tracing polynomial X-ray optics keeping the polynomial coefficients as symbolic and unevaluated until optimized.

\section {Monte-Carlo ray traces}
 \label{s:montecarlo}

\begin{figure}
\begin{center}
\begin{tabular}{c}
\includegraphics[width=6in]{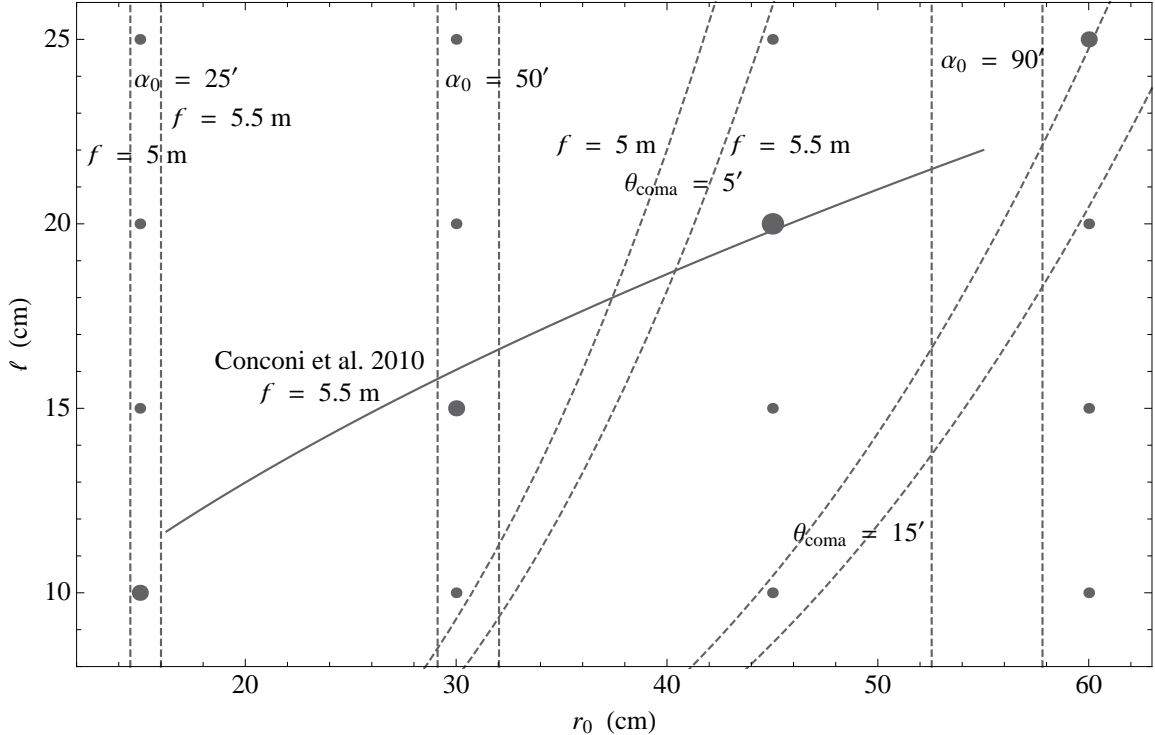}
\end{tabular}
\end{center}
\caption[example] 
{ \label{f:montecarlo} 
Mirror segment length $\ell_{s}$ vs. intersection radius $r_{0,s}$, with points showing locations in the $(r_{),s},\ell_{s})$ plane of Monte-Carlo ray traces with 50,000 incident rays for focal lengths of 5.5 m.
The largest dot shows the location of additional ray traces with 100,000 incident rays at a focal length of 5.5 m.
The largest and mid-size dots show locations of additional ray traces with 50,000 incident rays for focal lengths of 4.5, 5.0 and 6.0 m.
Note there are 2 segments per shell.
The solid line represents our reconstruction of the wide-field telescope design described in Ref. \citenum{Conconi10} using their design constraints.
The vertical dashed lines show constant values for nominal graze angles of 25, 50 and 90 arcmin at the intersection plane for chosen values of $r_0$.  The curved dashed lines show constant values of 5 and 15 arcmin for $\theta_{coma}$ [see \S\ref{s:nested}, Eq. (\ref{e:thetacoma})], in the $( \ell_{s}, r_{0,s})$ plane.}
\end{figure}

In order to explore the dependences of $\sigma^2(\theta,\phi)$ on mirror shell parameters, we carried out an extensive series of Monte-Carlo ray traces of single shell Wolter I optics for nominal focal lengths, $f$, of 5.5 m, mirror shell segment lengths (2 segments per shell), $\ell_{s}$, of 10, 15, 20 and 40 cm, and shell intersection radii, $r_{0,s}$, of 15, 30, 45 and 60 cm.
Here the subscript $s$ denotes a shell number.
Figure (\ref{f:montecarlo}) plots the locations of these ray traces in the $\ell_{s}$ vs. $r_{0,s}$ plane.
In most cases, the number of rays incident on the shell aperture was 50,000; for the point in Figure (\ref{f:montecarlo}) marked with the biggest dot the number of incident rays was 100,000.
The mid-size dots show locations of ray traces for focal lengths of 4.5, 5.0 and 6.0 m.
We used the results for $\sigma^2$ from these ray traces to devise trial analytic functions for representing $\sigma_{s}^2(\theta,\phi)$ as a function of the angles $\theta$, $\phi$, and of $f$, $\ell_{s}$ and $r_{0,s}$.

The solid curve in Figure (\ref{f:montecarlo}) shows our reconstruction of the relationship for $\ell_{s}$ vs. $r_{0,s}$ for the 3 telescope module, 82 mirror shell per module wide-field design described and discussed in Ref. \citenum{Conconi10}.
We carried out this reconstruction using the design constraints provided in their Table (2).
While certain of our assumptions may vary from theirs, in general we expect our reconstruction to be close to their actual design.

\section {Single mirror shell}
\label{s:single1}

For Monte-Carlo ray traces of a single mirror shell $s$, we define the geometric area, $A_{geom,s}$, as

\begin {equation}
 \label{e:aeff}
 A_{geom,s}(\theta) \ \equiv \ A_{inc,s} \ n_{s}(\theta) \ / \ n_{inc,s}(\theta),
\end {equation}

\noindent
where $A_{inc,s}$ is the entrance aperture for shell $s$, $n_{inc,s}$ the number of rays incident on that aperture, and $n_{s}$ the number of doubly reflected rays exiting the mirror shell.



On a detector tilted by an angle $\theta_{tilt}$ with one corner at the $( x, y )$ origin, but displaced along the optical axis by an amount $\delta z_{s}$ from the flat plane perpendicular to the optical axis at the nominal on-axis focus, the variance, or square of the RMS dispersion may be written in the form

\begin {equation}
 \label {e:rms}
 \sigma_{s}^2(\theta, \phi, \delta z_{s}, \theta_{tilt}) \ = \ a_{s} \ + \\ 2 \ b_{s} \ \delta z_{s} \ + \ c_{s} \ \delta z_{s}^2 \ + \ 2 \ d_{s} \ \tan{\theta_{tilt}} \ + \ 2 \ e_{s} \ \delta z_{s} \ \tan{\theta_{tilt}} \ + \ f_{s} \ \tan^2{\theta_{tilt}}.
\end {equation}

\noindent
Evaluating the merit function [Eq. (\ref{e:merit})] for this shell leads to

\begin {eqnarray}
 \label{e:merit1}
 M(\delta z_{s}, \theta_{tilt}) & = & a_{s,M} \ + \ 2 \ b_{s,M} \ \delta z_{s,M} \ + \ c_{s,M} \ \delta z_{s,M}^2 \nonumber \\
   &   &   \nonumber \\
   &   & + \ 2 \ d_{s,M} \ \tan{\theta_{tilt}} \ + \ 2 \ e_{s,M} \ \delta z_{s} \ \tan{\theta_{tilt}} \ + \ f_{s,M} \ \tan^2{\theta_{tilt}},
\end {eqnarray}

\noindent where the subscript $M$ denotes an average over the FOV like that in Eq. (\ref{e:merit}).
In order to carry out the integrals over the FOV, it is advantageous to have analytic forms for the coefficients $a_s$, $b_s$, $c_s$, $d_s$, $e_s$ and $f_s$.
Minimizing Eq. (\ref{e:rms}) with respect to $\delta z_s$ and $\tan{\theta_{tilt}}$, we find

\begin {equation}
 \label{e:minima1}
 \tan{\theta_{tilt}} \ =  \ \left( \frac{b_{s,M} \ e_{s,M} \ - \ c_{s,M} \ d_{s,M}}{c_{s,M} \ f_{s,M} \ - \ e_{s,M}^2} \right)
\end {equation}

\begin {equation}
 \label{e:minima2}
 \delta z_s \ = \ \left( \frac{d_{s,M} \ e_{s,M} \ - \ b_{s,M} \ f_{s,M}}{c_{s,M} \ f_{s,M} \ - \ e_{s,M}^2} \right)
\end {equation}

\noindent
Expressions (\ref{e:rms})---(\ref{e:minima2}) are general and applicable to any surface prescription for grazing incidence X-ray optics.

Below we use the notation:

\begin {equation}
 \label{e:notation1}
 \sum_{x,y} \ < ( x, y ) \left( \frac{k_{(x,y)}}{k_z} \right) > \ = \ < x \left( \frac{k_x}{k_z} \right) > \ + \ < y \left( \frac{k_y}{k_z} \right) >,
\end {equation}

\noindent and similarly for other combinations of terms.
Angle brackets around a quantity, $< q >$, denote an average over that quantity on an the focal surface.
Angle brackets with the subscript 0, $< q >_0$, denote an average in the flat plane perpendicular to the optical axis at the nominal on-axis focal position.
We find that the coefficients $a_s$, $b_s$, $c_s$, $d_s$, $e_s$ and $f_s$ can be expressed in terms of averages in that flat plane.
Assuming a detector in the first quadrant (both $x$ and $y$ positive), here we provide some examples of coefficient definitions along with the trial fitting functions that we find useful for representing the results of our Monte-Carlo ray traces:

\begin {equation}
 \label{e:acoeff1}
 \sigma_{s}^2(0,0) \ \equiv \ a_{s} \ \equiv \ \sum_{(x,y)} \ ( \ < (x,y)^2 >_{0,s} \ - \ < (x,y) >_{0,s}^2 \ ),
\end {equation}

\noindent
The behavior of $a_s$ as a function of $\theta$ is complicated by effects due to coma.
At present, we are working with the three trial fitting functions, $a_0(\theta)$, $a_1(\theta)$ and $a_2(\theta)$, for Wolter I optics.
We define a useful function, $g$, and then $a_0(\theta)$, $a_1(\theta)$ and $a_2(\theta)$:

\begin {equation}
 \label{e:gdefns}
 g(\theta, \ \zeta, \ \xi) \ = \ 1 \ + \ \zeta \ \tan{\theta} \ + \ \xi \ \tan^2{\theta} \nonumber \\
\end {equation}

\begin {eqnarray}
 \label{e:acoeff2}
 a_{fit}(\theta) \ = \ a_{coma}(\theta) \ + \ a_{m}(\theta), \ (m = 0, \ 1, \ 2) \ \ \ \ \ \ \ \ \ \ \ \ \ \ \ \ \ \ \ \ \ \ \ \ \ \ \ \ \ \ \ \ \ \ \ \nonumber  \\
   \nonumber \\
 a_{coma}(\theta) \ = \ ( \tan{4 \alpha_0} \ / \ 2 \ )^4 \ \tan^2{\theta} \ \ \ \ \ \ \ \ \ \ a_0(\theta) \ = \ ( \ 2 \ \mu_{a,0} \ \ell \ / \ \tan{4 \alpha_0} \ )^2 \ \tan^4{\theta} \ g(\theta, \ \zeta_{a}, \ \xi_{a}) \\
   \nonumber \\
 a_{1}(\theta) \ = \ a_{coma}(\theta) \ + \ a_0(\theta) \ \ \ \ \ \ \ \ \ \  a_{2}(\theta) \ = \ \left( \sqrt{a_{coma}(\theta)} \ + \ \sqrt{a_0(\theta)} \right)^2, \ \ \ \ \ \ \ \ \ \ \ \ \ \ \ \ \ \nonumber
\end {eqnarray}

\noindent
We note that $a_1$ defined here is equivalent to Eq. (2) in Ref. \citenum{LVS72}.
Additional examples of coefficient definitions and trial fitting functions applicable to Wolter I optics are:

\begin {eqnarray}
  \label{e:bcoeff1}
 b_{s} & \equiv & \sum_{(x,y)} \ \left[ < (x,y) \left( \frac{k_{(x,y)}}{k_z} \right) >_{0,s} \ - \ < (x,y) >_{0,s} < \left( \frac{k_{(x,y)}}{k_z} \right) >_{0,s} \right], \nonumber \\
   &   &   \\
 b_{fit}(\theta) & = & 2 \ \mu_b \ \ell \ \tan^2{\theta} \ g(\theta, \ \zeta_{b}, \ \xi_{b}), \nonumber
\end {eqnarray}

\begin {eqnarray}
  \label{e:ccoeff1}
 c_{s} & \equiv & \sum_{(x,y)} \ \left[ < \left( \frac{k_{(x,y)}}{k_z} \right)^2 >_{0,s} \ - \ < \left( \frac{k_{(x,y)}}{k_z} \right) >_{0,s}^2 \right], \nonumber \\
   &   &   \\
 c_{fit}(\theta) & = & [ \ \mu_c \ \tan{4 \alpha_0} \ g(\theta, \ \zeta_{c}, \ \xi_{c}) \ ]^2. \nonumber
\end {eqnarray}

\noindent
Definitions and trial fitting functions for $d_s$, $e_s$, and $f_s$ are lengthy, so for reasons of readability we provide them in Appendix \ref{a:def}.

\section {Nested mirror shells}
\label{s:nested}

Consider a set of $S$ nested telescopes.
We assume uniform illumination of the entrance aperture for the full array of telescopes.
The number of rays through the $s$-th telescope is $n_s$, and the total number of rays through the array of nested telescopes is

\begin {equation}
\label{e:N}
 N \ \equiv \ \sum_{s=1}^S n_s.
\end {equation}

\noindent
We designate the $( x, y, z )$ position coordinates on an arbitrary focal surface for the full array of the $k$-th ray through the $s$-th telescope by $( \ x_{s,k}, \ y_{s,k}, \ z_{s,k} \ )$.
We find that the total variance, or square of the RMS disperion, for the full set of nested shells may be written in the form

\begin {equation}
\label{e:sigma}
 \sigma^2 \ = \ \sigma_{1}^2 \ + \ \sigma_{2}^2,
\end {equation}

\noindent
where

\begin {equation}
\label{e:sigma1}
 \sigma_{1}^2 \ = \ \sum_{s=1}^S \ \left( \frac{n_s - 1}{N-1} \right) \ \sigma_{s}^2(\delta z_s, \theta_{tilt}), \nonumber \\
\end{equation}

\noindent
with $\sigma_{s}^2(\delta z_s, \theta_{tilt})$ given by Eq. (\ref{e:rms}).
The second term on the right-hand-side of Eq. (\ref{e:sigma}) is given by

\begin {equation}
\label{e:sigma2}
 \sigma_{2}^2 \ = \ \left(\frac{N}{N-1}\right) \ \sum_{(x,y,z)} \ \left[ \sum_{s=1}^S  \frac{n_s}{N} < (x,y,z)_s >^2 \ - \ \left( \sum_{s=1}^S \frac{n_s}{N} < (x,y,z)_s > \right)^2 \right],
\end {equation}

Eqs. (\ref{e:sigma})---(\ref{e:sigma2}) show that the variance, $\sigma^2$, for the full set of nested shells has two contributions.
The first, Eq. (\ref{e:sigma1}), is a sum over the variances for the individual telescopes, on the chosen focal surface, weighted essentially by their relative effective geometric areas $[ (n_s / N) \simeq A_{geom,s}(\theta) / \sum_s A_{geom,s}(\theta)]$.
The second, Eq. (\ref{e:sigma2}), is a sum over a kind of weighted variance of the means, $< (x,y,z)_s >$, for the individual telescopes on that focal surface.
This second contribution can be viewed as arising from the differences in the best focal surfaces for the individual mirror shells from that for the full set of nested shells (see Ref. [\citenum{Conconi10}]).
Expressions (\ref{e:sigma})---(\ref{e:sigma2}) are general and applicable to any surface prescription for grazing incidence X-ray optics.

For best performance, Eqs. (\ref{e:sigma})---(\ref{e:sigma2}) mean that the minimization of $\sigma_M^2$, and thus the optimization of the parameters $\theta_{tilt}$ and $(\delta z_s, s = 1, 2, 3, ... S)$, must be done simultaneously, rather than following Eqs. (\ref{e:minima1}) and (\ref{e:minima2}) for the individual shells.
In principle this can be done using matrix methods, although the number of linear equations involved is large for current wide field designs which approach 100 nested mirror shells (see Ref. [\citenum{Conconi10}]).

We have also shown a need for expressions for $< ( x, y, z )_s >$, and terms, $< ( x, y, z )_s >^2$ and cross terms $< ( x, y, z ) > < ( x, y , z) >$, that can then be derived to the appropriate order, for the individual shells.
We find to the appropriate order

 \begin {eqnarray}
 \label{e:means}
 < ( x, y ) >_s & = & a^{\prime}_{(x,y),s} \ + \ b^\prime_{(x,y),s} \ \delta z_{s} \ + \ d^\prime_{(x,y),s} \ \tan{\theta_{tilt}} \ + \ e^\prime_{(x,y),s} \delta z_{s} \ \tan{\theta_{tilt}} \ + \ f^\prime_{(x,y),s} \ \tan^2{\theta_{tilt}} \nonumber \\
   &   &   \\
 < z >_s & = & ( \ a^{\prime}_{x,s} \ + \ a^{\prime}_{y,s} \ ) \tan_{\theta{tilt}} \ + \ b^{\prime}_{x,s} \ + \ ( \ b^{\prime}_{y,s} \ ) \ \delta z_{s} \ \tan{\theta_{tilt}} \ + \ ( \ d^{\prime}_{x,s} \ + \ d^{\prime}_{y,s} \ ) \ \tan^2{\theta_{tilt}}. \nonumber
\end{eqnarray}

\noindent
We provide the definitions of $a^{\prime}_{(x,y),s}$, $b^{\prime}_{(x,y),s}$, $d^{\prime}_{(x,y),s}$, $e^{\prime}_{(x,y),s}$ and $f^{\prime}_{(x,y),s}$ in Appendix \ref{a:def}.
For use below, we define an angle, $\theta_{coma}$, at which $a_{coma}(\theta)$ and $a_0(\theta)$ [see Eq. (\ref{e:acoeff2})], with $\zeta_a = 0$ and $\xi_a = 0$, are equal:

\begin {equation}
 \label{e:thetacoma}
 \tan{\theta_{coma}} \ = \ \left( \frac{1}{8} \right) \left( \frac{f}{\ell} \right) \ \tan^3{4 \alpha_0}.
\end {equation}

\noindent
We have not yet finished devising analytic expressions for $e_{(x,y),s}^{\prime}$ and for $f_{(x,y),s}^{\prime}$, but here provide trial fitting functions for for $a_{(x,y),s}^{\prime}$, $b_{(x,y),s}^{\prime}$ and $d_{(x,y),s}^{\prime}$:

\begin {equation}
 \label{e:acoeffxy1}
 a^{\prime}_{(x,y),fit} \ = \ f \ ( \ 1 \ + \ \delta f_{(x,y),a} \ ) \ \tan{\theta} \ \left[ 1 \ + \ \left( \frac{3}{4} \right) \ \tan^2{\theta} \right] \ g(\theta, \ \zeta_{(x,y),a}, \ \xi_{(x,y),a} ),
\end {equation}

\begin {eqnarray}
 \label{e:bcoeffxy1}
 b^{\prime}_{(x,y),fit,0}(\theta, \phi) \ = \ - \ \mu_{(x,y),b} \ \tan{\theta} \ ( \cos{\phi}, \ \sin{\phi} ) \ g(\theta, \ \zeta_{(x,y),b}, \ \xi_{(x,y),b} ) \ \ \ \ \ \ \ \ \ \ \ \ \ \ \ \ \ \ \ \ \nonumber \\
   \nonumber \\
 k_{(x,y),coma}(\theta) \ = \ p_{(x,y),coma} \ \sin{[ \ \pi \ \zeta_{(x,y),coma} \ ( \theta \ / \theta_{coma} ) \ ] } \ \ \ \ \  \ \  \ \ \ \ \ \ \ \ \ \ \ \ \ \ \ \ \ \ \ \ \ \ \ \ \ \ \nonumber \\
   \\
 k_{(x,y),damp}(\theta) \ = \ \exp{ [ \ - \ \xi_{(x,y),coma} \ ( \theta / \theta_{coma} )^2 \ ] } \ \ \ \ \ \ \ \ \ \ \ \ \ \ \ \ \ \ \ \ \ \ \ \ \ \ \ \ \ \ \ \ \ \ \ \ \ \ \ \ \ \ \ \ \ \ \nonumber \\
   \nonumber \\
 b^{\prime}_{(x,y),fit}(\theta, \phi) \ = \ k_{(x,y),damp}(\theta) \  k_{(x,y),coma}(\theta) \ + \ ( 1 \ - \ k_{(x,y),damp}(\theta) ) \ [ \ q_{(x,y),coma} \ + \ b^{\prime}_{(x,y),fit,0}(\theta, \phi) \ ], \nonumber
\end {eqnarray}

\begin {eqnarray}
 \label{e:dcoeffxy1}
 d^{\prime}_{(x,y),fit}(\theta, \phi) & = & - \ f \ ( 1 \ + \ \delta f_{(x,y),d} ) ( 1 \ + \ 2 \ \tan^2{4 \alpha_0} ) \ \tan^2{\theta} \nonumber \\
   &   &   \\
   &   & \times \ ( \cos{\phi}, \ \sin{\phi} ) \ ( \cos{\phi} \ + \ \sin{\phi} ) \ g(\theta, \ \zeta_{(x,y),d}, \ \xi_{(x,y),d} ), \nonumber
\end {eqnarray}

\noindent In the future, we plan to provide a fuller account of our methods and results for ray tracing Wolter I optics.

\section {Notation for polynomial coefficients}
 \label{s:polynotation}

In the past, we have written sums over rays

\begin {equation}
 \label{e:sum}
 \lambda \ = \ \sum_{k=1}^{n} \lambda_{k},
\end {equation}

\noindent where $\lambda_{k}$ is some quantity such as position along an axis or a component of a direction vector for ray $k$, in the form of a second order expansion in the polynomial coefficients

\begin {eqnarray}
 \label{e:oldnotation}
 \lambda & = & \lambda_{0000} \ + \ u_{1} \ \lambda_{1000} \ + \ u_{2} \ \lambda_{0100} \ + \ u_{3} \ \lambda_{0010} \ + \ u_{4} \ \lambda_{0001} \nonumber \\
           &   & \nonumber \\
           &   & + \ u_{1}^2 \ \lambda_{2000} \ + \ u_{2}^2 \ \lambda_{0200} \ + \ u_{3}^2 \ \lambda_{0020} \ + \ u_{4}^2 \ \lambda_{0002} \nonumber \\
           &   & \\
           &   & + \ u_{1} \ u_{2} \ \lambda_{1100} \ + \ u_{1} \ u_{3} \ \lambda_{1010} \ + \ u_{1} \ u_{4} \ \lambda_{1001} \nonumber \\
           &   & \nonumber \\
           &   & + \ u_{2} \ u_{3} \ \lambda_{0110} \ + \ u_{2} \ u_{4} \ \lambda_{0101} \ + \ u_{3} \ u_{4} \ \lambda_{0011} \nonumber,
\end {eqnarray}

\noindent where $u_1$, $u_2$, $u_3$, and $u_4$ are the polynomial coefficients for the mirror shell.
The form Eq.~(\ref{e:oldnotation}) assumes the coefficients are small enough so that a second order expansion is valid.
The polynomial deviations from Wolter I optics required in the applications we have studied so far satisfy this criterion.

We define a polynomial coefficient vector 

\begin {equation}
 \label{e:polyvector}
 \vec{u} \ \equiv \ (u_{1}, u_{2}, u_{3}, u_{4}).
\end {equation}

\noindent We also define the scalars


 \begin {eqnarray}
 \label{e:scalars}
\lambda_{00} \ \equiv \ \lambda_{0000} & \lambda_{01} \ \equiv \ \lambda_{1000} & \lambda_{02} \ \equiv \ \lambda_{0100} \nonumber \\
   &   &  \nonumber \\
 \lambda_{03} \ \equiv \ \lambda_{0010} & \lambda_{04} \ \equiv \ \lambda_{0001} &
  \lambda_{11} \ \equiv \ \lambda_{2000} \nonumber \\
   &   &  \nonumber \\
 \lambda_{22} \ \equiv \ \lambda_{0200} & \lambda_{33} \ \equiv \ \lambda_{0030} & \lambda_{44} \ \equiv \ \lambda_{0002} \\
   &   &  \nonumber \\
  \lambda_{12} \ = \lambda_{21} \ \equiv \ \frac{1}{2}\lambda_{1100} & \lambda_{13} \ = \lambda_{31} \ \equiv \ \frac{1}{2}\lambda_{1010} & \lambda_{14} \ = \lambda_{41} \ \equiv \ \frac{1}{2}\lambda_{1001} \nonumber \\
   &   &  \nonumber \\
  \lambda_{23} \ = \lambda_{32} \ \equiv \ \frac{1}{2}\lambda_{0110} & \lambda_{24} \ = \lambda_{42} \ \equiv \ \frac{1}{2}\lambda_{0101} & \lambda_{34} \ = \lambda_{43} \ \equiv \ \frac{1}{2}\lambda_{0011} \nonumber,
 \end {eqnarray}


\noindent and the vectors

\begin {eqnarray}
 \label{e:vectors}
 \vec{\lambda_{0}} & \equiv & (\lambda_{01}, \lambda_{02}, \lambda_{03}, \lambda_{04}) \nonumber \\
   &   & \nonumber \\
 \vec{\lambda_{1}} & \equiv & (\lambda_{11}, \lambda_{12}, \lambda_{13}, \lambda_{14}) \nonumber \\
   &   & \nonumber \\
 \vec{\lambda_{2}} & \equiv & (\lambda_{21}, \lambda_{22}, \lambda_{23}, \lambda_{24}) \\
   &   & \nonumber \\
 \vec{\lambda_{3}} & \equiv & (\lambda_{31}, \lambda_{32}, \lambda_{33}, \lambda_{34}) \nonumber \\
   &   & \nonumber \\
 \vec{\lambda_{4}} & \equiv & (\lambda_{41}, \lambda_{42}, \lambda_{43}, \lambda_{44}). \nonumber
\end {eqnarray}

\noindent Finally we define a matrix with the row vectors $\vec{\lambda_{1}}$, $\vec{\lambda_{2}}$, $\vec{\lambda_{3}}$, and $\vec{\lambda_{4}}$:

\begin {equation}
 \label{e:matrix}
 \overline{\overline{\lambda}} \ \equiv \ \left(\begin {array}{c}
 \vec{\lambda_{1}} \\
 \vec{\lambda_{2}} \\
 \vec{\lambda_{3}} \\ 
 \vec{\lambda_{4}}\end{array}\right) \ = \ \left(\begin {array}{cccc}
 \lambda_{11} & \lambda_{12} & \lambda_{13} & \lambda_{14} \\
 \lambda_{21} & \lambda_{22} & \lambda_{23} & \lambda_{24} \\
 \lambda_{31} & \lambda_{32} & \lambda_{33} & \lambda_{34} \\
 \lambda_{41} & \lambda_{42} & \lambda_{43} & \lambda_{44}
\end{array}\right).
\end {equation}


\begin {equation}
 \label{e:newnotation}
 \lambda \ = \ \lambda_{00} \ + \ \vec{u} \cdot \vec{\lambda_{0}} \ + \ \vec{u} \cdot \overline{\overline{\lambda}} \cdot \vec{u}.
\end {equation}

\noindent In this notation, and since the matrix $\overline{\overline{\lambda}}$ is symmetric, derivatives of $\lambda$ with respect to the polynomial coefficients are

\begin {equation}
 \label{e:polyder1}
 \frac{\partial \lambda}{\partial u_{j}} \ = \ \lambda_{0j} \ + \ 2 \sum_{i=1}^4 \ \lambda_{ji} \ u_{i} \ = \ \lambda_{0j} \ + \ 2 \  \vec{u} \cdot \vec{\lambda_{j}}.
\end {equation}

\noindent For a set of nested shells, denoted by prefixes $s$ or $r$ (\eg, $\lambda_s)$, we note that

\begin {equation}
 \label{e:polyder2}
 \partial \left(\sum_{r=1}^S \lambda_r\right)/\partial u_{s,j} \ = \ \frac{\partial \lambda_s}{\partial u_{s,j}}.
\end {equation}


\section {Outer product of two vectors}
 \label{s:outer}

Consider the two vectors

\begin {eqnarray}
 \label{e:outerexample1}
 \vec{a} & \equiv & ( a_1, a_2, a_3, a_4 ) \nonumber \\
   &   &   \\
 \vec{b} & \equiv & ( b_1, b_2, b_3, b_4 ) \nonumber.
\end {eqnarray}

\noindent In linear algebra, the outer product of $\vec{b}$ with $\vec{a}$ is given by

\begin {equation}
 \label{e:outerexample2}
 \vec{b} \otimes \vec{a} \ \equiv \ \left(\begin {array}{cccc}
 b_1 \ a_1 & b_1 \ a_2 & b_1 \ a_3 & b_1 \ a_4 \\
 b_2 \ a_1 & b_2 \ a_2 & b_2 \ a_3 & b_2 \ a_4 \\
 b_3 \ a_1 & b_3 \ a_2 & b_3 \ a_3 & b_3 \ a_4 \\
 b_4 \ a_1 & b_4 \ a_2 & b_4 \ a_3 & b_4 \ a_4
\end{array}\right).
\end {equation}

\noindent The outer product is an essential tool in polynomial ray tracing algebra.

\section {Basic operations}
 \label{s:basic}

Consider the two polynomial objects

\begin {eqnarray}
 \label{e:addexample1}
 a & = & a_{00} \ + \ \vec{u} \cdot \vec{a_0} \ + \ \vec{u} \cdot \overline{\overline{a}} \cdot \vec{u} \nonumber \\
   &   &   \nonumber \\
 b & = & b_{00} \ + \ \vec{u} \cdot \vec{b_0} \ + \ \vec{u} \cdot \overline{\overline{b}} \cdot \vec{u} \\
   &   &   \nonumber \\
 \vec{u} & = & ( u_1, u_2, u_3, u_4 ). \nonumber
\end {eqnarray}

\noindent Treating $\vec{u}$ as small and expanding to second order in $\vec{u}$, we now specify how to carry out basic operations on $\vec{a}$ and $\vec{b}$.
{\bf Using these operations, it is possible to construct a Monte-Carlo ray trace code for polynomial X-ray optics with sufficiently small but unknown coefficients $\vec{u}$.
In principle, values for the coefficients can then be derived from a final ray bundle for any assumed merit function.}


In the case of the addition operation, we note

\begin {equation}
 \label{e:addexample2}
 a \ \pm \ b \ = (a_{00} \pm b_{00}) \ + \ \vec{u} \cdot (\vec{a_0} \pm \vec{b_0}) \ + \ \vec{u} \cdot (\overline{\overline{a}} \pm \overline{\overline{b}}) \cdot \vec{u}.
\end {equation}


The multiplication operation is more complicated.  We want to keep terms only to 2nd order in the polynomial coefficients $\vec{u}$.  To this order we find

\begin {equation}
 \label{e:multexample1}
 a \times b \ = \ a_{00} \ b_{00} \ + \ \vec{u} \cdot (a_{00} \ \vec{b_0} \ + \ b_{00} \ \vec{a_0} ) \ + \ \vec{u} \cdot (a_{00} \ \overline{\overline{b}} \ + \ b_{00} \ \overline{\overline{a}} \ + \ \vec{b_0} \otimes \vec{a_0}) \cdot \vec{u}.
\end {equation}

\noindent In particular, we note that

\begin {equation}
 \label{e:multexample2}
 (\vec{u} \cdot \vec{a_0}) \times (\vec{u} \cdot \vec{b_0}) \ = \ \vec{u} \cdot (\vec{b_0} \otimes \vec{a_0}) \cdot \vec{u}.
\end {equation}

\noindent We also need the square and the cube of a polynomial object to 2nd order in $\vec{u}$.  We find

\begin {eqnarray}
 \label{e:squarecube}
 a^2 & = & a_{00}^2 \ + \ \vec{u} \cdot (2 \ a_{00} \ \vec{a_0}) \ + \ \vec{u} \cdot ( 2 \ a_{00} \ \overline{\overline{a}} \ + \ \vec{a_0} \otimes \vec{a_0} ) \cdot \vec{u} \nonumber \\
   &   &   \\
 a^3 & = & a_{00}^3 \ + \ \vec{u} \cdot (3 \ a_{00}^2 \ \vec{a_0}) \ + \ \vec{u} \cdot ( 3 \ a_{00}^2 \ \overline{\overline{a}} \ + \ 3 \ a_{00} \ \vec{a_0} \otimes \vec{a_0} ) \cdot \vec{u}. \nonumber
\end {eqnarray}


For the division operation, to second order in $\vec{u}$, we find

\begin {eqnarray}
 \label{e:division}
 a / b & = & \left( \frac{a_{00}}{b_{00}} \right) \ \left\{ 1 \ + \ \vec{u} \cdot \left[ \left( \frac{\vec{a_0}}{a_{00}} \right) \ + \ \left( \frac{\vec{b_0}}{b_{00}} \right) \right] \right\} \nonumber \\
   &   &   \\
   &   & + \ \left( \frac{a_{00}}{b_{00}} \right) \ \left\{ \vec{u} \cdot \left[ \left( \frac{\overline{\overline{a}}}{a_{00}} \right) \ - \ \left( \frac{\overline{\overline{b}}}{b_{00}} \right) \ + \ \left( \frac{\vec{b_0}}{b_{00}} \right) \otimes \left( \frac{\vec{b_0}}{b_{00}} \right) \ - \ \left( \frac{\vec{b_0}}{b_{00}} \right) \otimes \left( \frac{\vec{a_0}}{a_{00}} \right) \right] \cdot \vec{u} \right\}. \nonumber
\end {eqnarray}


In the case of the square root operation, to second order in $\vec{u}$, we find

\begin {eqnarray}
 \label{e:sqroot}
 \sqrt{a} & = & \sqrt{a_{00}} \ \left\{ 1 \ + \ \frac{1}{2} \ \vec{u} \cdot \left( \frac{\vec{a_0}}{a_{00}} \right) \ + \ \frac{1}{2} \ \vec{u} \cdot \left( \frac{\overline{\overline{a}}}{a_{00}} \right) \cdot \vec{u} \ - \ \frac{1}{8} \ \vec{u} \cdot \left[ \left( \frac{\vec{a_0}}{a_{00}} \right) \otimes \left( \frac{\vec{a_0}}{a_{00}} \right) \right] \cdot \vec{u}. \right\} \\
   &   &   \nonumber \\
 1 \ / \ \sqrt{a} & = & 1 \ / \ \sqrt{a_{00}} \ \left\{ 1 \ - \ \frac{1}{2} \ \vec{u} \cdot \left( \frac{\vec{a_0}}{a_{00}} \right) \ - \ \frac{1}{2} \ \vec{u} \cdot \left( \frac{\overline{\overline{a}}}{a_{00}} \right) \cdot \vec{u} \ + \ \frac{3}{8} \ \vec{u} \cdot \left[ \left( \frac{\vec{a_0}}{a_{00}} \right) \otimes \left( \frac{\vec{a_0}}{a_{00}} \right) \right] \cdot \vec{u}. \right\} 
\end {eqnarray}

\noindent We also find

\begin {eqnarray}
 \label{e:boverroota}
 a \ / \ \sqrt{b} & = & \frac{a_{00}}{\sqrt{b_{00}}} \ \left\{ 1 \ + \ \vec{u} \cdot \left[ \left( \frac{\vec{a_0}}{a_{00}} \right) \ - \ \frac{1}{2} \ \left( \frac{\vec{b_0}}{b_{00}} \right) \right] \right\} \nonumber \\
   &   &   \\
   &   & + \ \frac{a_{00}}{\sqrt{b_{00}}} \ \left\{ \vec{u} \cdot \left[ \left(  \frac{\overline{\overline{a}}}{a_{00}} \right) \ - \ \frac{1}{2} \ \left(  \frac{\overline{\overline{b}}}{b_{00}} \right) \ - \ \frac{1}{2} \ \left( \frac{\vec{b_0}}{b_{00}} \right) \otimes \left( \frac{\vec{a_0}}{a_{00}} \right) \ + \ \frac{3}{8} \ \left( \frac{\vec{b_0}}{b_{00}} \right) \otimes \left( \frac{\vec{b_0}}{b_{00}} \right) \right] \cdot \vec{u} \right\}. \nonumber
\end {eqnarray}

\noindent Eq.~(\ref{e:boverroota}) is especially useful for normalizing unit vectors such as ray direction vectors or normal vectors to surfaces.

\section {Propagation of rays}
 \label{s:rayprop}

The new $x$ and $y$ coordinates of rays propagated from axial position $z_1$ to $z_2$ according to

\begin {eqnarray}
 \label{e:xyznew}
 x_2 & = & x_1 \ + \ k_{x1} \ t \nonumber \\
   &   &   \nonumber \\
 y_2 & = & y_1 \ + \ k_{y1} \ t \\
   &   &   \nonumber \\
 z_2 & = & z_1 \ + \ k_{z1} \ t. \nonumber
\end {eqnarray}

\noindent Since $z_1$ and $k_{z1}$ are known, the parametric variable $t$ can be expressed in terms of those quantities and the value for $z_2$.
In polynomial notation, we have

\begin {eqnarray}
 \label{e:tnew}
 t & = & t_{00} \ + \ \vec{u} \cdot \vec{t} \ + \ \vec{u} \cdot \overline{\overline{t}} \cdot \vec{u} \ = \ \frac{z_2 - z_1}{k_{z1}} \nonumber \\
   &   &   \\
   &   & \left[ \ \left( z_{2,00} - z_{1,00} \right) + \vec{u} \cdot \left( \vec{z_2} - \vec{z_1} \right) + \vec{u} \cdot \left( \overline{\overline{z_2}} - \overline{\overline{z_1}} \right) \cdot \vec{u} \ \right] \ / \ \left( k_{z1,00} + \vec{u} \cdot \vec{k_{z1}} + \vec{u} \cdot \overline{\overline{k_{z1}}} \cdot \vec{u} \right). \nonumber
\end {eqnarray}

\noindent The rule for division, Eq.~(\ref{e:division}), shows how to evaluate the polynomial components of $t$ namely ($t_{00}$, $\vec{t}$, and $\overline{\overline{t}}$) from Eq.~(\ref{e:tnew}) given the known polynomial components of $z_1$, $z_2$, and $\vec{k_1}$.
Then, also given $x_1$, $y_1$, $k_{x1}$, and $k_{y1}$ in the form of polynomial objects, we can apply the basic operations of addition and multiplication defined in \S\ref{s:basic} to Eq.~(\ref{e:xyznew}) to find $x_2$ and $y_2$ in the form of polynomial objects also.  {\bf This means that knowing the polynomial components of the initial position and direction vector, we can compute the polynomial components of the final position vector without knowing numerical values for the polynomial coefficients $\vec{u}$.}

\section {Surface prescriptions for polynomial X-ray optics}
 \label{s:poly}

We consider mirror prescriptions for the primary mirrors of the form

\begin {eqnarray}
 \label{e:surface}
 r_{s}^2(z) & = & r_{0,s}^2 \ \left[ 1 \ + \ 2 \ A_{s} \ (z/r_{0,s}) \ + B_{s} \ (z/r_{0,s})^2 \ + \ u_{a,s} \ (z/r_{0,s})^2 \ + \ u_{b,s} \ (z/r_{0,s})^3 \right] \nonumber \\
   &   &   \\
   & = & r_{0,s}^2 \ \left[ 1 \ + \ 2 \ A_{s} \ (z/r_{0,s}) \ + B_{s} \ (z/r_{0,s})^2\ + \ (z/r_{0,s})^2 \ \vec{u_{s}} \cdot \vec{\zeta_{0,s}} \right] . \nonumber
\end {eqnarray}

\noindent For the primary (P) and secondary (S) mirror segments, we have

\begin {equation}
 \label{e:segments}
 \begin {array}{llll}
 P: & A_{s} = \tan{\alpha_{0,s}}, & B_{s} = 0, & \zeta_{0,s,P} = \left( 1, \left(\frac{z}{r_{0,s}} \right), 0, 0 \right) \\
   &   &   &   \\
 S: & A_{s} = \tan{3\alpha_{0,s}}, & B_{s} = h(\alpha_{0,s}) \tan^2{3 \alpha_{0,s}}, & \zeta_{0,s,S} = \left( 0, 0, 1, \left(\frac{z}{r_{0,s}} \right) \right).
\end {array}
\end {equation}

\noindent Here 

\begin {eqnarray}
\label{e:alpha0}
 \alpha_{0,s} & = & \left( \frac{1}{4} \right) \ \tan^{-1}{\left( \frac{r_{0,s}}{f} \right)} \nonumber \\
   &   &   \\
 h(\alpha_{0,s}) & = & 1-1/[1+2 \cos{(2 \alpha_{0,s})}]^2, \nonumber
\end {eqnarray}

\noindent and $\vec{u} = ( u_{a,s,P}, u_{b,s,P}, u_{a,s,S}, u_{b,s,S} )$.
The notation and methods introduced here are, in principle, readily extended to additional terms in the mirror presciptions [\eg, proportional to $(z/r_{0,s})^4$, $(z/r_{0,s})^5$, etc.].

\section {Ray tracing polynomial X-ray optics}
\label{s:polyraytracing}

In order to trace rays through X-ray optics, one needs to:  (1) populate the entrances aperture with rays (both in position and direction);  (2)  calculate intersections with mirror segment surfaces;  (3) calculate the unit normals to the surfaces at those intersections, including deviations due to non-ideal surfaces; (4) determine the direction of the reflected ray; and (5) take account of obstruction by the next innermost mirror shell.
Using the tools outlined above, all these tasks can be accomplished, to sufficent accuracy, for polynomial X-ray optics as long as the polynomial coefficients are sufficiently small.
In the future, we plan to provide more details on this method and its results.

\section {Concluding remarks}
\label{s:remarks}

The ultimate goal of our work is to reduce the complexity of design procedures for nested grazing incidence X-ray telescopes, specifically those with Wolter I and polynomial designs.
In this paper, we have:

\begin {enumerate}
\item Described our use of Monte-Carlo ray traces to devise trial analytic formulae for the coefficients of terms in the expression [Eq. (\ref{e:rms})] for the spatial variance of rays from a  point source on an arbitrary focal surface for a single Wolter I mirror shell.
Our adopted merit function [Eq. (\ref{e:merit})] can then be minimized to provide the best displacement of the mirror shell along the optical axis and the best value for the detector tilt angle.
\item Shown that for a set of nested mirror shells, the spatial variance on an arbitrary focal surface is a sum of two terms.
The first [Eq. (\ref{e:sigma1})] is a sum over the variances of the individual shells evaluated on that focal surface, weighted by their relative effective areas.
The second [Eq. (\ref{e:sigma2})] is a sum over a kind of variance for the mean positions of rays from the individual shells on that focal surface.
The existence of this second term means that it is necessary to optimize parameters such as mirror shell displacement along the optical axis, detector tilt angle, and polynomial coefficients simultaneously for all mirror shells, rather than individually.
\item In \S\ref{s:polynotation}---\S\ref{s:polyraytracing}, introduced notation and mathematical tools for ray tracing polynomial optics leaving the polynomial coefficients in symbolic form.
In principle, this simplifies the design procedure by reducing the required number of Monte-Carlo ray traces, and permitting determination of numerical values for the polynomial coefficients through the solution of a large number of linear equations derived from minimization of the merit function.
\end {enumerate}

Our future plans are to continue these studies, refining the trial analytic functions for Wolter I optics, implementing a polynomial optic ray trace code using the tools described in this paper, and hopefully providing a less complex means for the optimization of wide-field X-ray telescope designs.


\acknowledgments     

We thank R. Giacconi, S. S. Murray, G. Pareschi, and all the members of the WFXT team for many interesting and helpful discussions and ideas.
We carry out all our X-ray optics ray trace work in the symbolic mathematics system {\sl Mathematica$^{\copyright}$}\cite{Wolfram03}, which makes much of our work easier, more accurate, and less tedious.

\bibliography{wfxt_100611}   
\bibliographystyle{spiebib}   

\clearpage


\appendix

\section {Additional coefficient expressions}
 \label{a:def}

We assume a detector in the first quadrant (both $x$ and $y$ positive).
First we provide definitions for the coefficients $d_s$, $e_s$ and $f_s$ (see \S\ref{s:single1}):

\begin {equation}
  \label{e:dcoeff1}
 d_{s} \ \equiv \ \sum_{(x,y)} \ \left[ < (x,y) \left( \frac{k_{(x,y)}}{k_z} \right) ( x + y ) >_{0,s} \ - \ < (x,y) >_{0,s} < \left( \frac{k_{(x,y)}}{k_z} \right) ( x + y ) >_{0,s}  \right],
\end {equation}

\begin {eqnarray}
  \label{e:ecoeff1}
 e_{s} & \equiv & \sum_{(x,y)} \ \left[ < \left( \frac{k_{(x,y)}}{k_z} \right)^2 ( x + y ) >_{0,s} \ - \ < \left( \frac{k_{(x,y)}}{k_z} \right) >_{0,s} < \left( \frac{k_{(x,y)}}{k_z} \right) ( x + y ) >_{0,s} \right] \nonumber \\
   &   &   \\
   &   & + \ \sum_{(x,y)} \ \left[ < (x,y) \left( \frac{k_{(x,y)}}{k_z} \right) \left( \frac{k_x+k_y}{k_z} \right) >_{0,s} \ - \ < (x,y) >_{0,s} < \left( \frac{k_{(x,y)}}{k_z} \right) \left( \frac{k_x+k_y}{k_z} \right) >_{0,s} \right], \nonumber
\end {eqnarray}

\begin {equation}
 \label{e:fcoeff}
 f_{s} \ = \ f_{xy,s} \ + \ f_{z,s},
\end {equation}

\begin {eqnarray}
  \label{e:fxycoeff1}
 f_{xy,s} & \equiv & \sum_{(x,y)} \ \left[ < ( x + y )^2 \left( \frac{k_{(x,y)}}{k_z} \right)^2 >_{0,s} \ - \ < ( x + y ) \left( \frac{k_{(x,y)}}{k_z} \right) >_{0,s}^2 \right] \nonumber \\
   &   &   \\
   &   & + \ 2 \ \sum_{(x,y)} \ \left[ < (x,y) ( x + y ) \left( \frac{k_{(x,y)}}{k_z} \right) \left( \frac{ k_x + k_y }{k_z} \right) >_{0,s} \ - \ < (x,y) >_{0,s} < ( x + y ) \left( \frac{k_{(x,y)}}{k_z} \right) \left( \frac{ k_x + k_y }{k_z} \right) >_{0,s} \right], \nonumber
\end {eqnarray}

\begin {equation}
  \label{e:fzcoeff1}
 f_{z,s} \ \equiv \ < ( x + y )^2 >_0 \ - \ < x + y >_0^2.
\end {equation}

\noindent
We typically find $f_{z,s} \ \ll \ f_{xy,s}$.
We define

\begin {eqnarray}
 \label{e:ghdefns}
 g(\theta, \ \zeta, \ \xi) & \equiv & 1 \ + \ \zeta \ \tan{\theta} \ + \ \xi \ \tan^2{\theta} \nonumber \\
   &   &   \nonumber \\
 h_{0}(\phi, \ \eta) & \equiv & 1 \ + \ \eta \ \sin{2 \phi} \\
   &   &   \nonumber \\
 h(\theta, \phi, \ \zeta, \ \xi, \ \eta_{0}, \ \eta_{\zeta}, \ \eta_{\xi} ) & \equiv & h_{0}(\phi, \ \eta_{0}) \ + \ \zeta \ \tan{\theta} \ h_{0}(\phi, \ \eta_{\zeta}) \ + \ \xi \ \tan^2{\theta} \ h_{0}(\phi, \ \eta_{\xi}). \nonumber
\end {eqnarray}

\noindent
Note that $h(\theta, \phi, \ \zeta, \ \xi, \ 0, \ 0, \ 0) \ = \ g(\theta, \phi, \ \xi)$.
Now we provide trial fitting functions for  the coefficients $d_s$, $e_s$ and $f_s$:

\begin {equation}
 \label{e:dcoeff2}
 d_{fit}(\theta, \phi) \ = \ 2 \ \mu_d \ f \ \ell \ \tan^3{\theta} \ h( \theta, \phi, \ \zeta_{d}, \ \xi_{d}, \ \eta_{0,d}, \ \eta_{\zeta,d}, \ \eta_{\xi,d} ),
\end {equation}


\begin {equation}
 \label{e:ecoeff2}
 e_{fit}(\theta, \phi) \ = \ \mu_e \ f \ \tan^2{4 \alpha_0} \ \tan{\theta} \ h( \theta, \phi, \ \zeta_{e}, \ \xi_{e}, \ \eta_{0,e}, \ \eta_{\zeta,e}, \ \eta_{\xi,e} ),
\end {equation}


\begin {equation}
 \label{e:fxycoeff2}
  f_{xy,fit}(\theta,\phi) \ = \ 2 \ \mu_{f,xy} \ ( \ f \ \tan{4 \alpha_0} \ \tan{\theta} \ )^2 \ h( \theta, \phi, \ \zeta_{f,xy}, \ \xi_{f,xy}, \ \eta_{0,f,xy}, \ \eta_{\zeta,f,xy}, \ \eta_{\xi,f,xy} ),
\end {equation}

\begin {equation}
 \label{e:fzcoeff2}
 f_{z,fit}(\theta,\phi) \ = \ \mu_{f,z} \ \tan^3{\theta} \ h( \theta, \phi, \ \zeta_{f,z}, \ \xi_{e}, \ \eta_{0,f,z}, \ \eta_{\zeta,f,z}, \ \eta_{\xi,f,z} ).
\end {equation}

We now provide the definitions of $a^{\prime}_{(x,y),s}$, $b^{\prime}_{(x,y),s}$, $d^{\prime}_{(x,y),s}$, $e^{\prime}_{(x,y),s}$ and $f^{\prime}_{(x,y),s}$ (see \S\ref{s:nested}):

\begin {eqnarray}
 \label{e:abdefcoeffxy1}
 a^{\prime}_{(x,y),s} \ = \ < ( x, y ) >_{0,s} \nonumber \\
   &   \nonumber \\
 b^{\prime}_{(x,y),s} \ = \ < \left( \frac{ k_{(x,y)}}{k_z} \right) >_{0,s} &  d^{\prime}_{(x,y),s} \ = \ < \left( \frac{ k_{(x,y)}}{k_z} \right) (x + y ) >_{0,s},\\
   &   \nonumber \\
 e^{\prime}_{(x,y),s} \ = \ < \left( \frac{ k_{(x,y)}}{k_z} \right) \left( \frac{k_x + k_y}{k_z} \right) >_{0,s}, & \ \ \ \ \ f^{\prime}_{(x,y),s} \ = \ < \left( \frac{ k_{(x,y)}}{k_z} \right) ( x + y ) \left( \frac{k_x + k_y}{k_z} \right) >_{0,s}. \nonumber
\end {eqnarray}








\end{document}